\documentclass[onecolumn]{aastex631}

\usepackage{amsmath}
\usepackage{subfigure}
\usepackage{booktabs}
\usepackage[hang,flushmargin]{footmisc}
\usepackage{multirow}
\usepackage{threeparttable}
\usepackage{hyperref}

\begin{document}

\title{{On the Fast-radio-burst-associated X-ray Bursts: Inverse Compton Scattering of Radio Photons by an Extreme Pair Flow During Magnetosphere Activities}}

\correspondingauthor{Yuan-Pei Yang (ypyang@ynu.edu.cn); Zi-Gao Dai (daizg@ustc.edu.cn)}

\author[0000-0002-5225-637X]{Yue Wu}
\affiliation{School of Astronomy and Space Science, Nanjing University, Nanjing 210023, China}

\author[0000-0001-6374-8313]{Yuan-Pei Yang}
\affiliation{South-Western Institute for Astronomy Research, Yunnan University, Kunming, Yunnan 650500, China}
\affiliation{Purple Mountain Observatory, Chinese Academy of Sciences, Nanjing 210023, China}

\author[0000-0003-4157-7714]{Fa-Yin Wang}
\affiliation{School of Astronomy and Space Science, Nanjing University, Nanjing 210023, China}
\affiliation{Key Laboratory of Modern Astronomy and Astrophysics (Nanjing University) Ministry of Education, Nanjing 210023, China}

\author[0000-0002-7835-8585]{Zi-Gao Dai}
\affiliation{Department of Astronomy, School of Physical Sciences, University of Science and Technology of China, Hefei 230026, China}

\begin{abstract}

The Galactic fast radio burst (FRB) FRB 200428 was associated with a short X-ray burst (XRB) from the magnetar SGR J1935+2154 during one of its active phases. This FRB-associated XRB exhibits distinct properties compared to other typical XRBs, including a significantly higher cutoff energy and a steeper power-law index. Its recovered X-ray light curve shows a multiple-peak structure, with the time of arrival offset from that of the FRB. These unique features imply a special physical link between the FRB and X-ray emissions. In 2022 October, a similar FRB-XRB association was detected from the same source. In this paper, we propose a model in which the observed spectral and temporal features of the associated XRBs can be attributed to the inverse Compton scattering (ICS) of FRB photons by an extreme pair flow around the light cylinder, with a bulk Lorentz factor of $\Gamma\sim10$ and a power-law distribution in the comoving frame, characterized by a typical Lorentz factor $\gamma^\prime_\mathrm{m}\sim5\times 10^4$. This extreme pair flow could originate from the compression of a transient pulse of $\sim10^{40}-10^{41}\mathrm{erg\,s^{-1}}$ and the acceleration through magnetic reconnection in the current sheet during magnetar activity. The Doppler-boosted ICS spectra and the arrival time shifts in such a scenario can well explain the observed features of the FRB 200428-associated XRB and can also account for another associated event in 2022.

\end{abstract}


\section{Introduction} \label{sec:intro}

Fast radio bursts (FRBs) are mysterious bright radio transients with millisecond durations and extremely high brightness temperatures \citep{XiaoDi21,Lyubarsky21,Petroff22,ZhangBing23}. 
So far, more than 800 FRB sources have been discovered, of which more than 70 exhibit repeating properties, with the repetitions ranging from twice to several thousand times (see the CHIME/FRB Public Database \footnote{\url{https://www.chime-frb.ca/} by the CHIME/FRB Collaboration.} and Blinkverse \footnote{\url{http://blinkverse.alkaidos.cn/} by the Zhejiang Lab.}). 
Some models have been proposed to explain the physical origin of FRBs. Based on the locations of the emission regions, these models can be generally classified into three main categories \citep{ZhangBing20a,ZhangBing23,XiaoDi21}: 
1) “close-in” or “pulsar-like” models, in which the emission is inside the neutron star magnetosphere \citep{Katz16,Kumar17,Kumar22,YangYP18,YangYP21,YangYP23a,DaiZG20,Kumar20,LuWB20,Wang2020,WangFY2022,WangWY22,ZhangBing22,YangYP23b,LiuZN23}; 2) “far-away” or “GRB-like” models, which invoke relativistic shocks situated far outside of the magnetosphere \citep{Lyubarsky14,Beloborodov17,Beloborodov20,Metzger19,Margalit20,Yamasaki22,Wada23}; 
and 3) the “intermediate-field” model, in which FRBs are generated near the light cylinder \citep{Lyubarsky20,Mahlmann22,WangJS23}. 

\begin{figure}[b]
\includegraphics[width=1\textwidth]{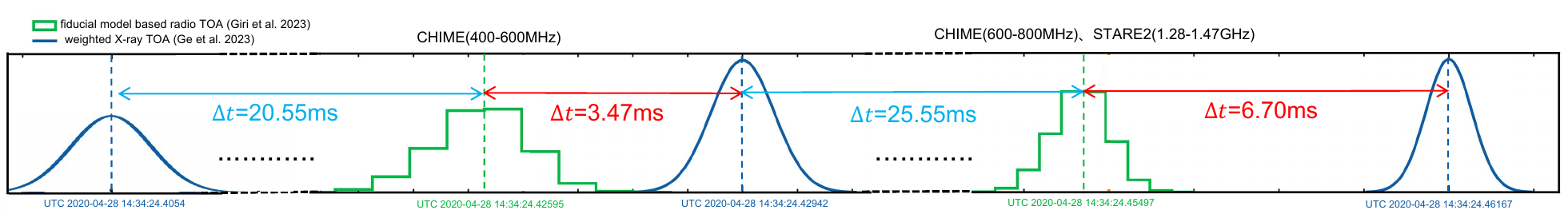}
\caption{{TOA and interval of the radio peaks of FRB 200428 and $\sim$millisecond X-ray peaks of the FRB-associated XRB (two “main” peaks and a “subpeak”) \citep{GeMY23,Giri23}. The green and blue dashed vertical lines denote the refined estimates for the TOAs of the radio and X-ray peaks, respectively. All the timestamps are corrected to an infinite frequency and referenced to the geocentric system. The $\sim0.2\,\mathrm{s}$ wide bump is not shown, and the time axis is not drawn to scale.} }
\label{Fig1}
\end{figure}

Although both the physical origin and the radiation mechanism of FRBs are still not well understood, magnetars have been confirmed as an origin of at least some FRBs. The Galactic FRB, FRB 200428, detected by CHIME \citep{CHIME20} and STARE2 \citep{Bochenek20}, was found to be associated with a bright X-ray burst (XRB) from the magnetar SGR J1935+2154 during one of its active phases \citep{Mereghetti20,LiCK21,Ridnaia21,Tavani21}. Its light curve is characterized by a $\sim0.2\,\mathrm{s}$ wide bump and multiple $\sim$millisecond bright peaks, with the times of arrival (TOAs) of these peaks showing time shifts of a few milliseconds relative to that of the FRB (as shown in Figure \ref{Fig1}) \citep{GeMY23,Giri23}. The XRB has a steeper power-law index and a much higher cutoff energy than other typical XRBs \citep{Younes21}, highlighting its unique properties and implying a special physical link between FRB 200428 and its X-ray counterpart. 

Some models involving magnetars have been proposed to interpret the basic features of the FRB-associated XRB, such as high-energy radiation processes, including the release of magnetic energy and resonant Compton scattering in the magnetosphere \citep{Yamasaki20,Yamasaki22,YangYP21}, an expanding fireball along open magnetic field lines \citep{Ioka20,Wada23,Wada24}, synchrotron radiation from internal shocks \citep{LiuZN23}, relativistic outflows \citep{Yamasaki22}, plasmoid ejection \citep{YuanYJ20,YuanYJ22}, and multiple Compton scatterings after QED magnetic reconnection \citep{XieYu23}. However, simultaneously accounting for both the spectral and temporal features, as well as their connections to FRBs, with minimal reliance on artificial external assumptions (the so-called ``Occam's Razor principle''), remains a great challenge. In addition, the reason why other XRBs lack these distinct properties is still poorly understood, although some studies have suggested that FRBs might have beaming radiation \citep{LinL20,ZhangBing21,ChenConnery23}. Consequently, the physical origin of the unique FRB-associated XRB remains to be further investigated.

Recently, in 2022 October, SGR J1935+2154 entered a new active episode, with multiple short XRBs detected over several days \citep{LiCK22,Palm22,Younes22a,HuCP24,Ibrahim24}. 
Similar to the active phase in 2020, during this episode, CHIME and the Green Bank Telescope (GBT) simultaneously detected an FRB-like radio burst with multiple radio peaks (denoted FRB 221014, for simplicity; \citealp{Maan22,Giri23}), which was also associated with an X-ray counterpart observed by GECAM and Konus-Wind. The peak energy fitted by an exponential cutoff model is $\sim40\, \mathrm{keV}$, which is also slightly harder than other typical bursts in this episode \citep{Frederiks22,WangCW22}.

In this paper, we propose a model in which both the hard cutoff energy and the TOA shifts of the X-ray peaks are explained by the inverse Compton scattering (ICS) of FRB photons by an extreme ephemeral relativistic pair flow near the light cylinder during a magnetar activity phase (where “ephemeral” means it lasts for only tens of milliseconds). Here, we refer to our model simply as the FRB-outflow ICS model. 
{And this pair flow might originate from the compression of a transient low-frequency pulse of $\sim10^{40}-10^{41}\mathrm{erg\,s^{-1}}$ and acceleration through magnetic reconnection in the current sheet during magnetar activity. }

This paper is organized as follows.
In Section \ref{sec:model}, we first present the basic concepts and assumptions of the model, discuss a possible formation scenario for the extreme pair flow, estimate its key parameters, and derive the resulting ICS spectrum and time shifts of the X-ray peaks.
In Section \ref{sec:fitting}, we apply this FRB-outflow ICS model to explain the observed temporal and spectral features of the FRB-200428-associated XRB and also discuss its application to the recent X-ray counterpart of FRB 221014. 
In Section \ref{sec:Discussion}, we provide some predictions and discuss the underlying assumptions and limitations of our model.
The conclusions are presented in Section \ref{sec:Conclusions}. The convention $Q_x = Q/10^{x}$ in cgs units is used throughout this paper.

~

\section{ICS of Radio Photons by an extreme pair flow}\label{sec:model}
\subsection{Model Formulation}
\begin{figure}[b]
\centering
\includegraphics[width=0.85 \textwidth]{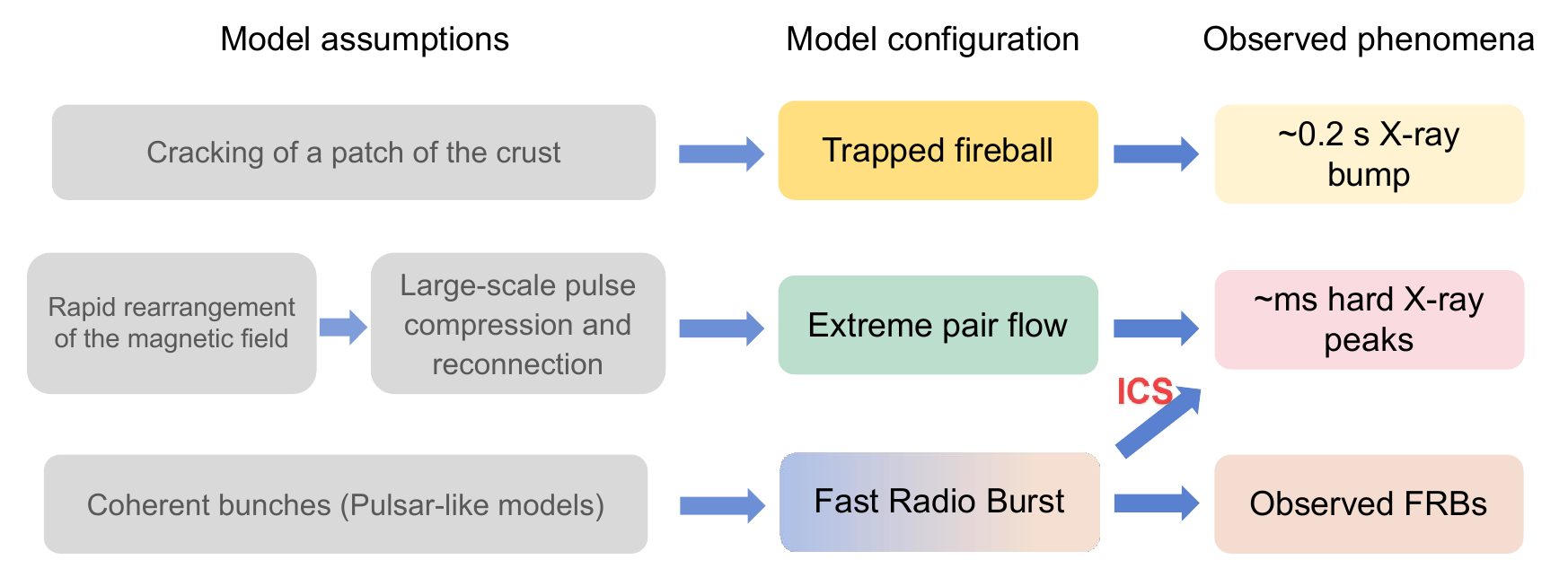}
\caption{{A flowchart showing the core concepts and assumptions of the model.}}\label{flowchart}
\end{figure}

{In our model, we assume the $\sim$0.2s wide bump and the multiple $\sim$millisecond bright X-ray peaks of the FRB 200428-associated XRB originate from distinct physical processes. 
The former (the wide bump) is attributed to the mechanisms commonly invoked for typical XRBs (e.g., \citealt{Thompson95,Lyubarsky02}).
In this work, we focus primarily on explaining the latter (the multiple $\sim$millisecond peaks), which exhibit much harder spectra and arrival time shifts \citep{Giri23}. 
We propose that these peaks arise from the ICS of the FRB photons by an extreme pair flow. 
The core concepts and assumptions of our model are outlined below and illustrated in the flowchart of Figure \ref{flowchart}):}

(1) The generation of the $\sim$0.2s X-ray bump. As in other typical short bursts of soft gamma repeaters (SGRs), the sudden cracking of a patch of the crust induced by magnetic stresses would create a confined photon pair plasma bubble, referred to as a “trapped fireball”. The spectrum of this trapped fireball is typically described as a modified blackbody spectrum, characterized by a superposition of Planckian spectra with $T < T_\mathrm{eff} \sim 10 \mathrm{keV}$ radiated from each layer of the E-mode photosphere, since the optical depth for E-mode photons depends on the frequency and becomes extremely large for O-mode photons \citep{Thompson95,Lyubarsky02,Yamasaki20}. This model is consistent with the observations of most typical XRBs.

(2) The formation of an extreme pair flow. The violent evolution of the magnetic field in the outer magnetosphere may lead to a rapid rearrangement, potentially generating a large-scale low-frequency pulse, with a luminosity of $\sim10^{40}-10^{41}\,\mathrm{erg\,s^{-1}}$ and opening the magnetosphere (the pulse produced in this region can propagate out without significant decay, with its duration roughly estimated as $\lesssim R_\mathrm{open}/c\sim 0.1 R_\mathrm{L}/c\sim 50\,\mathrm{ms}$). As the pulse propagates outward, it would interact with the magnetar wind, compressing and accelerating the plasma to a bulk Lorentz factor $\Gamma$ \citep{Lyubarsky20,WangJS23}. Violent magnetic reconnection is expected to be initiated in the current sheet, which could significantly dissipate the magnetic energy and energize the particles to a characteristic Lorentz factor $\gamma_\mathrm{m}^\prime$ in the comoving frame. We will discuss this process and estimate the typical value in detail in Section \ref{sec:formation}.

\begin{figure}
\centering
\subfigure[\label{Fig3a}]{
\includegraphics[width=0.65 \textwidth]{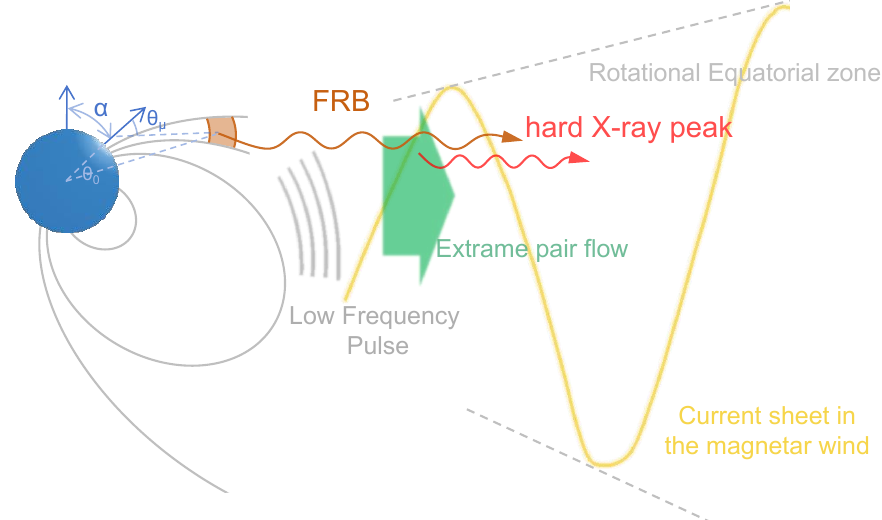}}
\
\subfigure[\label{Fig3b}]{
\includegraphics[width=0.75 \textwidth]{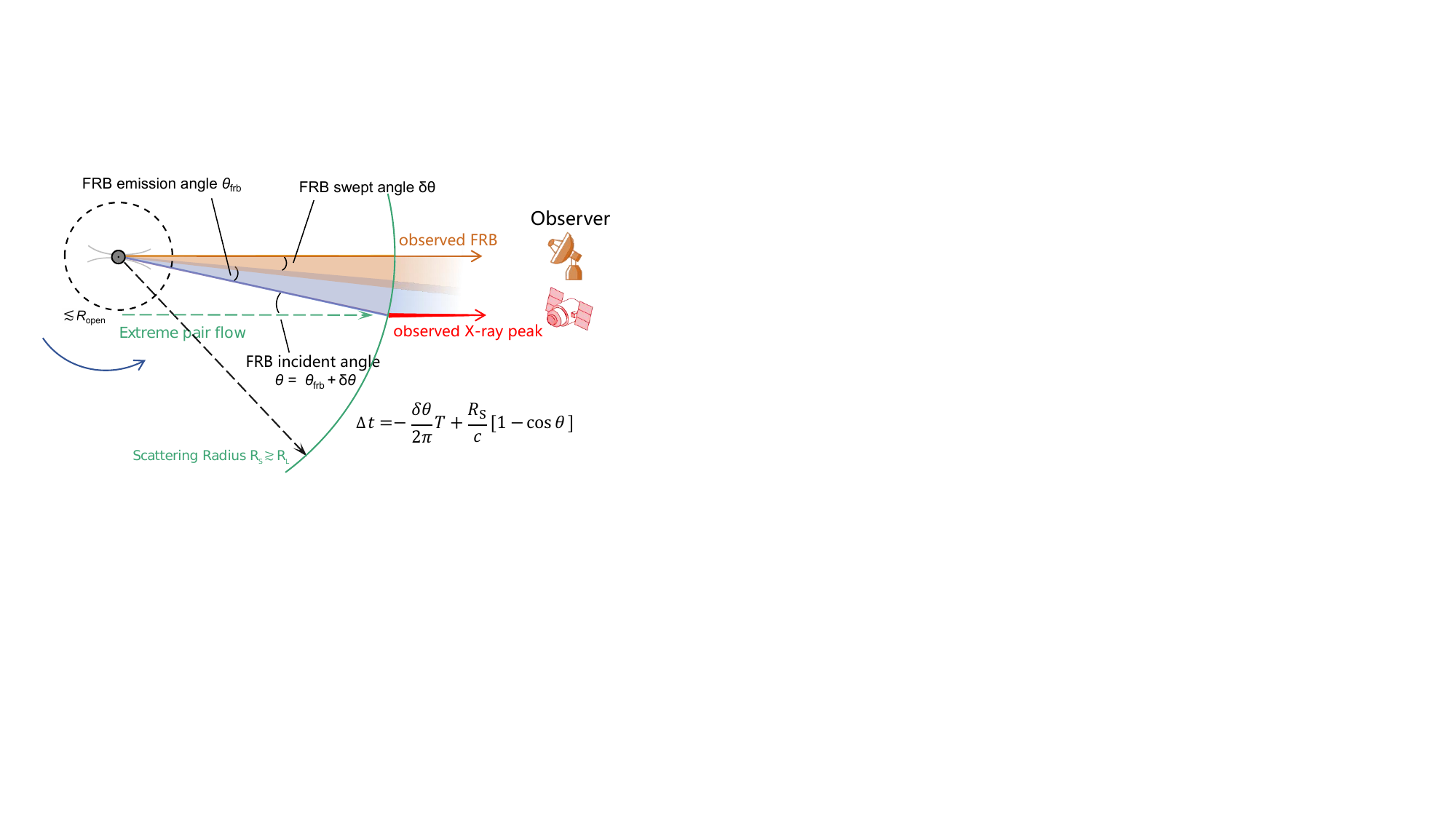} }
\caption{{Schematic configuration of the ICS of FRB photons by an extreme pair flow. 
An extreme pair flow is characterized by a bulk Lorentz factor of $\Gamma\sim10$, and the relativistic electrons in the comoving frame follow a power-law distribution with a typical local Lorentz factor of $\gamma^\prime_\mathrm{m}\sim 5\times10^4$. 
This extreme pair flow could originate from the compression by a transient pulse of $\sim10^{40-41}\mathrm{erg\,s^{-1}}$ and acceleration through magnetic reconnection in the current sheet during magnetar activity (Section \ref{sec:formation}).} 
(a) Side-view schematic configuration. The FRB is emitted in the magnetosphere, with an emission direction tangential to the local magnetic field line at the emission point, with the angle between the emission direction and the magnetic axis being denoted by $\theta_\mu$. 
The large-scale low-frequency pulse (gray thick line) from the rapid rearrangement of the magnetic field compresses the plasma and pushes it away. Violent reconnection is expected to be initiated in the current sheet, which energize the particles to a Lorentz $\gamma^\prime_\mathrm{m}$, composing the extreme pair flow (green thick line). 
The inclination angle $\alpha$ between the rotation and magnetic axes is expected to be $\gtrsim40^{\circ}$, to ensure the FRB emission pass through the wind zone where the extreme pair flow energy is likely to be predominantly concentrated.
(b) Top-view schematic configuration. The bulk motion of the flow slightly deviates from purely radial directions, with the azimuthal angle $\theta\lesssim R_\mathrm{open}/r$ (green dashed line). 
An FRB is triggered and emitted, with a small emission cone of opening angle $\theta_\mathrm{frb}$ (blue shaded area). 
Before pointing to the observer, radio photons from the magnetosphere would be scattered by relativistic electrons along the LOS, predominantly at the radius around $R_\mathrm{S}\gtrsim R_\mathrm{L}$, producing the hard-X-ray peak (red solid line). Subsequently, as the magnetar rotates with a small angle $\mathrm{\delta} \theta$, the FRB emission cone reaches the LOS, producing the observed FRB (brown shaded area and brown solid line). The time shift of the X-ray peak relative to the FRB peak $\Delta t$ arises from two competing contributions: (1) the phase difference, $-\mathrm{\delta} \theta T /2\pi$, resulting from the upscattered X-ray photons emitted at an earlier rotational phase; and (2) the path difference, $R_\mathrm{S}(1 - \cos \theta)/c$, caused by the additional distance traveled by the upscattered X-ray photons. A positive $\Delta t$ indicates a delayed X-ray peak, whereas a negative $\Delta t$ corresponds to an X-ray precursor. The schematic configuration is not drawn to scale.
}\label{Fig2}
\end{figure}

(3) Assumptions on the magnetar and the origin of the FRB. In this work, we assume that the FRB is generated in the magnetosphere, just after the pulse, with a small emission cone of opening angle $\theta_\mathrm{frb}$, consistent with pulsar-like models, where coherent radio radiation is concentrated along open field lines. We further assume that the FRB emission cone corotates with the magnetar and persists for several to tens of milliseconds, defining the intrinsic emission duration. Additionally, we assume that the magnetar is an oblique rotator with a large inclination angle $\alpha$ between the rotation and magnetic axes. As illustrated in Figure \ref{Fig3a} (from the side view), while most energy of the extreme pair flow may be concentrated in the rotational equatorial zone, the FRB photons could pass through this region and be upscattered, provided the magnetar has a large inclination angle. We will further discuss and constrain this angle in Section \ref{sec:Discussion}.

(4) The generation of the $\sim$millisecond hard-X-ray peaks. Thus, as shown in the top-view schematic, Figure \ref{Fig3b}, an FRB ignited before pointing to the observer (blue shaded area) would be upscattered by the relativistic electrons in the extreme flow, generating the time-shifted $\sim$millisecond hard-X-ray peaks. Subsequently, as the magnetar rotates with a small angle $\delta\theta$, the FRB emission cone reaches the line of sight (LOS), producing the observed FRB (brown shaded area). 
We next show that the spectrum and arrival time of the resulting X-ray emission produced via ICS primarily depend on the scattering location, the Lorentz factor of the electrons, the opening angle of the FRB emission cone, and the magnetar's rotation. Based on the FRB-outflow ICS model, the spectral hardness and arrival-time-shift features of the FRB 200428- and FRB 221014-associated hard-X-ray peaks could be well explained, regardless of whether these peaks precede or lag behind the corresponding FRB. 

\subsection{Characteristics of an extreme pair flow } \label{sec:formation}
Magnetar activity appears to be complex and diverse, and some previous studies have proposed various scenarios from different perspectives.
\citet{Younes23} and \citet{HuCP24} invoked a strong ephemeral wind to explain the rapid spindown observed in SGR J1935+2154, with a luminosity of $\sim10^{40}\,\mathrm{erg\,s^{-1}}$ and a duration of $\sim 10\, \mathrm{h}$, which could effectively carry away an amount of angular momentum during the antiglitch or the recovery phase between two large glitches.
In addition, it has been proposed that XRBs may be accompanied by some strong outflows confined within a small solid angle, due to the constraint of the magnetic field \citep{Ioka20,YuanYJ20,Wada23}. 
Some simulation studies have suggested that a powerful wind outflow during magnetar activity can open magnetic field lines and create an extended large-scale current sheet \citep{YuanYJ20,YuanYJ22,Mahlmann23}. 
Moreover, \cite{Lyubarsky20} proposed that a millisecond-duration, impulsive, high-luminosity pulse of $\sim10^{47}\,\mathrm{erg\,s^{-1}}$ could be generated in the outer magnetosphere and further give rise to fast magnetosonic (FMS) waves through the merging of magnetic islands near the light cylinder. These FMS waves are then converted into FRBs, providing a potential FRB origin mechanism.
A local 2D particle-in-cell simulation by \citep{Mahlmann22} has further demonstrated that this scenario can generate coherent gigahertz emission with luminosities sufficient for extragalactic FRBs and reproduce the frequency drifts and nanosecond substructures. However, the Galactic event FRB 200428 with a lower luminosity, does not fall within the suitable luminosity and frequency range of this scenario. In general, the complex and multifaceted nature of magnetar activity and its connection to the observation still require further investigation.

In this paper, we propose one possible scenario for the formation of an extreme pair flow. 
As described in the previous subsection, we assume that the rapid evolution of the outer magnetosphere generates a large-scale low-frequency pulse and opens the magnetosphere.
As it propagates outward, it would interact with the magnetar wind, compressing the plasma and pushing it forward, with the corresponding bulk Lorentz factor relative to the wind being given by \citep{Lyubarsky20,Lyubarsky21,WangJS23}
\begin{equation}\label{eq:Gamma}
        \Gamma'=\frac{1}{2}\sqrt{\frac{B_\mathrm{pulse}}{B_\mathrm{bg}}}=\frac{1}{2}\left(\frac{2LR_\mathrm{open}^4}{B_\mathrm{ns}^2 R_\mathrm{ns}^6c}\right)^{1/4},
\end{equation}
where $B_\mathrm{pulse}=(L/c)^{1/2}/R$ is the amplitude of the pulse and $L\sim10^{40-41}\,\mathrm{erg\,s^{-1}}$ is the total isotropic luminosity. (Since we mainly consider the region near the light cylinder, where the wind Lorentz factor remains low \footnote{Considering the wind Lorentz factor grows linearly up to the FMS point \citep{Beskin98}.}, we can use $\Gamma'$ as the Lorentz factor in the lab frame, i.e. $\Gamma\gtrsim\Gamma'$.) The background magnetic field is given by $B_\mathrm{bg}=B_\mathrm{open}R_\mathrm{open}/R$, representing the field in the opened-up region outside the critical radius $R_\mathrm{open}\lesssim R_\mathrm{L}$, at which the particle kinetic energy density begins to exceed the dipole magnetic field energy density:
\begin{equation}
     R_\mathrm{open}\simeq \left( \frac{B_\mathrm{ns}^2 R_\mathrm{ns}^6c}{2L_\mathrm{p}}\right)^{1/4},
\end{equation}
where $L_\mathrm{p}$ is the particle kinetic luminosity (prior to further acceleration), which can be written as $L_\mathrm{p}\simeq 4 \pi r^2 c\Gamma n m_\mathrm{e} c^2$. 
Here, $n$ denotes the number density of the pair flow, $B_\mathrm{ns}$ is the surface magnetic field strength, and $R_\mathrm{ns}$ is the magnetar radius. 
Then, violent magnetic reconnection is expected to occur within the current sheet, which significantly dissipates magnetic energy and energizes particles in the comoving frame.
The reconnection process has been studied analytically in some previous works (e.g., \citealp{Lyubarskii96,Petri12,Uzdensky14,Lyubarsky20}, etc.). 
For convenience, we analyze the physical process in the comoving frame that moves along with the current sheet. 
The pressure balance between the external magnetic field and the relativistic pairs implies
\begin{equation}
    n^{\prime} \gamma_\mathrm{m}^{\prime} \mathrm{m_e} \mathrm{c}^2 =\frac{B_\mathrm{pulse}^{\prime 2}}{8 \pi}=\frac{L}{4 \pi R_\mathrm{open}^2 \mathrm{c} \Gamma^2} ,
\end{equation}
where the prime denotes quantities measured in the plasma comoving frame and $\gamma_\mathrm{m}^\prime$ is the characteristic Lorentz factor of the energized particles. Assuming that the energy released via reconnection is balanced by the synchrotron cooling, the energy conservation implies
\begin{equation}
    \eta_{\mathrm{rec}} \frac{c}{4 \pi} B_\mathrm{pulse}^{\prime 2}=\frac{4}{9} n^{\prime} \mathrm{c} \mathrm{r_e}^2 B_\mathrm{pulse}^{\prime 2} \gamma_{\mathrm{m}}^{\prime 2} \Delta^{\prime},
\end{equation}
where $\eta_{\mathrm{rec}}$ is the reconnection rate, $r_\mathrm{e}$ is the electron radius, and $\Delta^{\prime}$ is the width of the layer, which is assumed to be $\xi$ times the Larmor radius and is given by
\begin{equation}\label{eq:layer}
    \Delta^{\prime}=\xi \frac{\gamma_\mathrm{m}^{\prime} \mathrm{m}_{\mathrm{e}} \mathrm{c}^2}{\mathrm{e} B_\mathrm{pulse}^{\prime}}.
\end{equation}
Thus, solving the above equations, the bulk Lorentz factor, the characteristic Lorentz factor, the critical radius, and the number density can be obtained by
\begin{equation}
\begin{aligned}
\Gamma &=\left(\frac{3^8 B_{\mathrm{ns}}^2 \mathrm{R}_{\mathrm{ns}}^6 \mathrm{e}^4 \mathrm{c}^3 \eta_{\mathrm{rec}}^4}{2^{35} r_e^8 \xi^4 L^3}\right)^{1 / 24} \sim 10\, \eta_{\mathrm{rec},-2}^{1 / 6} \xi_1^{-1 / 6} L_{40}^{-1 / 8}
\\
\gamma_{\mathrm{m}}^{\prime} &=
\frac{1}{2}\left(\frac{3^4{\mathrm{e}}^2 c^{3/2} \eta_\mathrm{rec}^2 \Gamma^2 B_\mathrm{p} R_\mathrm{NS}^3}{r_\mathrm{e}^4 \xi^2 L^{3/2}}  \right)^{2/7}
\sim 5\times 10^4 \, \eta_{\mathrm{rec},-2}^{2 /3} \xi_{1}^{-2 / 3} L_{40}^{-1 / 2}  , 
\\
R_\mathrm{open}&=\frac{1}{\sqrt{2}} \left(\frac{9\mathrm{e}2c^{5/2}\eta_\mathrm{rec} \Gamma B_\mathrm{p}^4 R_\mathrm{NS}^{12}}{r_\mathrm{e}^2 \xi L^{5/2}}  \right)^{1/7} 
\sim0.5 R_\mathrm{L} \, \eta_{\mathrm{rec},-2}^{1 / 6} \xi_{1}^{-1 / 6} L_{40}^{-3 / 8}  ,
\\
\dot{N}^{\prime}&=4 \pi r^2 \mathrm{c} n^{\prime}=\frac{ L}{ \Gamma^2 \gamma_{\mathrm{c}}^{\prime} \mathrm{m}_{\mathrm{e}} \mathrm{c}^2}
\sim 10^{39} \mathrm{~s}^{-1} \, \eta_{\mathrm{rec},-2}^{-1} \xi_{1} L_{40}^{7 / 4} . \label{eq:gamma}
\end{aligned}
\end{equation} 
Considering that the real environment during the magnetar activity phase is highly complicated and diverse, we simply assume that the electron/positron pairs follow a power-law distribution around the characteristic energy in the comoving frame, with sharp low- and high-energy cutoffs. 
Since the X-ray flux of the ICS process is primarily contributed by electron/positron pairs, we focus on the energy carried by them, denoted as $L_\pm$. If additional baryonic components are present and extract a fraction of the energy, $L_\pm$ would be smaller than $L$, thereby suppressing the ICS emission. In the following, we assume that most of the energy is carried by the pairs and neglect the difference between $L_\pm$ and $L$. 
Furthermore, the total power could be lower than the typical isotropic-equivalent value mentioned above. 

In addition to the conversion of energy, the angular momentum of the electromagnetic field is also transferred to the particles, causing the bulk motion to deviate slightly from a purely radial direction, with a small angle $\theta=v_\phi/c\lesssim R_\mathrm{open}/r$ \citep{Bogovalov00,Aharonian03,Aharonian12}. Here, we treat $\theta$ as a free parameter, which can be constrained by the observed spectrum and time shift. 

Moreover, during the reconnection, the typical photon energy of synchrotron radiation is given by
\begin{equation}
    \varepsilon_{\mathrm{sym,c}}=\Gamma \frac{0.2 \mathrm{e} \gamma_{\mathrm{c}}^{\prime 2} \mathrm{B}_{\mathrm{pulse }}^{\prime}}{2 \pi m_{\mathrm{e}} c} \sim 500 \mathrm{keV} \eta_{\mathrm {rec, -2}}^{7 / 6} \xi_1^{-7 / 6} L_{40}^{-1 / 8},
\end{equation}
and its luminosity satisfies $L_\mathrm{syn} < \eta_\mathrm{rec} L$, corresponding to a photon flux $7\times 10^{-5}\mathrm{ph\, cm^{-2} s^{-1} keV^{-1}}$, which is lower than the tail of the FRB-outflow ICS spectrum. Therefore, the X-ray emission from magnetic reconnection can be neglected in comparison with the ICS emission.
In addition, the magnetosonic waves generated by the merging of magnetic islands are expected to exhibit a characteristic frequency \citep{Lyubarsky20,WangJS2020},
\begin{equation}
    \nu= \frac{2 \Gamma }{2 \pi }\frac{c}{\kappa \Delta' } \sim 10^5\,
    \mathrm{Hz}\,\kappa_1^{-1}\eta_{\mathrm {rec, -2}}^{5 / 6} \xi_1^{-1 / 6} L_{40}^{11 / 8},
\end{equation}
where the size of the magnetic islands is $\kappa$ times that of the current sheet. The resulting frequency is
far below the gigahertz band and well beneath the detection thresholds of current telescopes. 
This is primarily due to the low energy of the burst from the galactic magnetar, as also discussed in \citet{WangJS2020}.

\subsection{The spectrum and time shift of the X-Ray peak }

As shown in the schematic configuration of Figure \ref{Fig2}, an FRB is triggered and emitted with a narrow emission cone of opening angle $\theta_\mathrm{frb}$ before rotating into the LOS (blue shaded area). This radiation can be upscattered by relativistic pairs in the extreme pair flow at a radius $r$ with a small incident angle $\theta$ (for simplicity, we assume the LOS lies on the equatorial rotation plane). 
The ICS of FRB photons by relativistic pairs in the flow can generate significant high-energy emission, potentially contributing to the hard-X-ray peaks of the FRB-associated XRB. 
Due to the relativistic bulk motion of the pair flow, the scattered emission is subject to Doppler boosting. The resulting X-ray spectrum produced by the ICS of FRB photons can be expressed as
\begin{align} \label{eq:DopplerEIC}
\frac{d N_{\mathrm{X}}}{d E_{\mathrm{X}} d S d t}= \frac{1}{4 \pi D_{\mathrm{L}}^2} \iiint \mathcal{D}^{-1}_\mathrm{obs} \frac{d N^{\prime}_{\mathrm{e}}(r)}{d \gamma^{\prime}d l^{\prime}} c\frac{d \sigma^{\prime}\left[\gamma^{\prime}, \theta^{\prime}, E^{\prime}_{\mathrm{X}}, E^{\prime}_{\mathrm{R}}\right]}{d E_\mathrm{X}^{\prime}} \frac{d n^{\prime}_{\mathrm{R}}}{d E^{\prime}_{\mathrm{R}} } \mathrm{~d} E^{\prime}_{\mathrm{R}} d \gamma^{\prime} d l,
\end{align} 
where $\mathcal{D}_\mathrm{obs} =1/\Gamma(1-\beta\cos\theta_\mathrm{obs}) $ is the Doppler factor and $\theta_\mathrm{obs}$ is the angle between the LOS and the bulk moving direction of the pair flow. The luminosity distance of the source is taken to be $D_\mathrm{L}=6.6\ \mathrm{kpc}$ \citep{ZhouP20}. And the integration over $d l$ is performed along the LOS.
Quantities in the plasma comoving frame are denoted by primes. The energies of the upscattered X-ray and incident radio photons in this frame are given by $ E^{\prime}_{\mathrm{X}} = E_{\mathrm{X}} / \mathcal{D}_\mathrm{obs} $ and $ E^{\prime}_{\mathrm{R}} = E_{\mathrm{R}} / \mathcal{D}_{*} $, respectively, where $\mathcal{D}_{*} =1/\Gamma(1-\beta\cos\theta_{*}) $ and $\theta_{*}$ is the incident angle between the direction of the radio photons and the bulk motion of the pair flow. 
The particle energy is assumed to follow a power-law distribution, as $dN'_\mathrm{e}/d\gamma'=K\gamma'^{-p}$, and $dN'_\mathrm{e}/dl^{\prime}$ denotes the number density of particles per length in the comoving frame, which can be obtained by
\begin{align}
L=c\Gamma^2 \int_{\gamma'_{\min}}^{\gamma'_{\max}}\frac{dN'_\mathrm{e}}{d\gamma'dl^{\prime}}\gamma'm_\mathrm{e}c^2d\gamma'
\end{align}  
The differential cross section for anisotropic ICS is given by \citep{Aharonian81,Fan08}
\begin{align}\label{eq:crosssection}
\frac{d \sigma'\left[\gamma', \theta', E'_{\mathrm{X}}, E'_{\mathrm{R}}\right]}{d E_\mathrm{X}'}= 
\frac{3 \sigma_{\mathrm{T}}}{4 {\gamma^{\prime}_{\mathrm{e}}}^2 E'_{\mathrm{R}} } \left[1+\frac{\xi^2}{2(1-\xi)}-\frac{2 \xi}{b_\theta(1-\xi)}+\frac{2 \xi^2}{b_\theta^2(1-\xi)^2}\right]
\end{align}
where $\xi \equiv E^{\prime}_{\mathrm{X}} /\left(\gamma^{\prime} m_{\mathrm{e}} c^2 \right)$ and $b_\theta=2\left(1-\cos \theta_{\mathrm{sc}}^{\prime}\right) \gamma^{\prime} E^{\prime}_\mathrm{R} /\left(m_{\mathrm{e}} c^2\right)$. Here, $\theta^{\prime}_\mathrm{SC}$ is the scattering angle between incident radio photons and scattered photons in the comoving frame, which is related to that in the observer frame by $1-\cos{\theta_\mathrm{sc}^{\prime}=\mathcal{D}_\mathrm{obs}\mathcal{D}_\mathrm{*}(1-\cos\theta)}$ \citep{Dubus10}. 

The ICS process is highly sensitive to the distance from the scattering site to the magnetar, since both the number densities of the FRB photons and relativistic electrons decrease as the square of the radius, and the incident angle $\theta\lesssim R_\mathrm{open}/r$ also becomes smaller at a larger radius. As a result, the radiation flux drops rapidly with radius, indicating that the emission is predominantly contributed by the region closest to the magnetar--that is, around $R_\mathrm{S}$. As the extreme pair flow may originate from the compression of a transient pulse and the acceleration through magnetic reconnection in the current sheet, the acceleration radius of the extreme pair flow would be approximately greater than the light cylinder $R_\mathrm{L}$. Then, the minimum scattering radius is defined as $R_\mathrm{S}\equiv \max\{R_\mathrm{L},R_\mathrm{s}\} \gtrsim R_\mathrm{L}$, where $R_\mathrm{s}$ represents the intersection radius between the FRB radiation and the LOS pair flow, which indicates the starting direction of the FRB radiation. 
As the magnetar rotates, the FRB emission cone sweeps a small angle of $\mathrm{\delta}\theta$, reaching the LOS and making the FRB observable (brown shaded area). As shown in Figure \ref{Fig2}, the geometrical relationship among the incident angle $\theta$ (also the scattering angle), the opening angle of the FRB emission cone  $\theta_\mathrm{frb}$, and the FRB swept angle $\mathrm{\delta}\theta$ can be expressed as 
\begin{align}    \label{eq:theta}
    \theta = \theta_\mathrm{frb}+\mathrm{\delta}\theta \lesssim R_\mathrm{open}/R_\mathrm{S}.
\end{align}  

Furthermore, because of the extremely high luminosity of the FRB emission, an FRB generated within a magnetar magnetosphere would go through the regime of “strong waves,” where the electric field amplitude of the FRB pulse $E_\mathrm{w}$ exceeds the background magnetic field $B_\mathrm{bg}$. In this case, the strength parameter $a \equiv eE_\mathrm{w}/m_\mathrm{e}c\omega_\mathrm{frb} > \omega_\mathrm{B}/\omega_\mathrm{frb} > 1$, where $\omega_\mathrm{B}$ is the cyclotron frequency and $\omega_\mathrm{frb}$ is the FRB frequency.
In this region, the Lorentz force contributed by the waves and the relativistic motion of electrons must be considered, leading to a significantly enhanced scattering cross section \citep{YangYP20,Beloborodov21,QuYH22,HuangYC24}. 
However, if the angle between the FRB propagation direction and the background magnetic field, as well as the opening angle of the radiation cone, are small, the interaction between the FRB pulse and the plasma would reduce substantially \citep{QuYH22,HuangYC24}. 
This allows FRBs generated in the magnetar magnetosphere to escape unimpeded. 
Moreover, the numerical result of the scattering cross section depends on the ratio $\omega_\mathrm{B}/\omega_\mathrm{frb}$. 
It decreases significantly, approaching the order of the Thomson cross section $\sigma_\mathrm{T}$, as $\omega_\mathrm{B}$ drops below $\omega_\mathrm{frb}$\citep{QuYH22}. 
The critical radius at which $\omega_\mathrm{B} = \omega_\mathrm{frb}$ can be defined as $R_\omega \equiv [e B_\mathrm{ns} R_\mathrm{ns}^3 / ( m_\mathrm{e} c \omega_\mathrm{frb} ) ]^{1/3} \simeq 0.6 R_\mathrm{L}$, beyond which the cross section approaches $\sigma_\mathrm{T}$. 
Therefore, for the scattering radius considered in this work ($R_\mathrm{S}$ approximates several $R_\mathrm{L}$), the nonlinear effect of the strong wave could be neglected.
On the other hand, the effect of scattering on the FRB itself could also be ignored, since
the optical depth $\tau(E_\mathrm{R}) \sim 10^{-8}\ll1$, for typical parameter values.
Based on the observation results of FRB 200428 and its associated XRB, and considering the immense photon flux of FRBs, it is expected that only a small fraction of the radio photons upscattered into X-rays to account for the observed XRB flux.

The time shift between the observed X-ray peak and the FRB peak is contributed by two competing components: (1) the phase difference--the up-scattered X-ray photons are emitted at an early rotational phase, introducing an advancement of $\mathrm{\delta}\theta \, T/2\pi$, where $T=3.24\,\mathrm{s}$ is the rotation period; (2) the path difference--the X-ray photons travel an additional length of $R_\mathrm{S}\left[1- \cos \theta \right]$ compared to the FRB photons, resulting in a delay.
The former is an advanced effect, while the latter introduces a time lag. These two factors compete with each other, resulting in either an advance or a delay of the X-ray peak. The time shift can be expressed as 
\begin{align}
    \Delta t = - \frac{\mathrm{\delta}\theta}{2\pi} T + \frac{R_\mathrm{S}}{c}\left(1- \cos \theta \right),
\label{eq:deltat}
\end{align}
where $\Delta t>0$ means that the X-ray peak lags behind the FRB, while $\Delta t<0$ represents an advanced X-ray peak.
This relation implies that a less-beamed FRB is more likely to produce a delayed X-ray peak, whereas a highly beamed FRB may be accompanied by an X-ray precursor.

\section{Explanation of observations }\label{sec:fitting}
In this section, we first briefly summarize the data analysis of FRB 200428 and its associated X-ray counterparts.
Then, we fit both the spectrum and the peak time shift and constrain the properties of the extreme pair flow and the FRB radiation, considering different combinations, in which the X-ray peaks either delay or advance the FRB. We also discuss the recent X-ray counterpart to FRB 221014, showing that the FRB-outflow ICS model could well account for the different spectral indices and advanced time. 

\subsection{Observational properties of FRB and associated XRB} 

The Galactic FRB 200428 consists of two sub-bursts. The first peak was detected by CHIME, showing an apparent spectral cutoff at 600 MHz, while the second peak extends from the CHIME 500-800 MHz band up to the STARE2 1.28-1.47 GHz band \citep{Bochenek20,CHIME20}. Based on the observed narrow spectrum feature, in this work, we simply assume that the FRB spectrum can be modeled by a Gaussian function:
\begin{equation}
 L_{\mathrm{frb},\nu}(\nu) = L_\nu(\nu_ \mathrm{p})\exp\left[ -\frac{1}{2} \left( \frac{\nu-\nu_\mathrm{p}}{\delta\nu}\right) \right],
\end{equation}   \label{eq:Gaussian}
where $\nu_ \mathrm{p} $ and $\delta \nu$ denote the peak frequency and characteristic width of FRB 200428, respectively. Assuming $\delta \nu / \nu_\mathrm{p} = 0.2$, we can get $\nu_\mathrm{p,1}=500 \ \mathrm{MHz}$, $\delta \nu_\mathrm{1}=100 \ \mathrm{MHz}$ for the first peak, and $\nu_\mathrm{p,2} = 1085 \ \mathrm{MHz}$, $\delta \nu_\mathrm{2} = 217 \ \mathrm{MHz}$ for the second peak.

The spectrum of the FRB-associated XRB exhibits unique properties, with a steeper power-law index and a much higher cutoff energy $\sim80 \mathrm{keV}$, than that of the other typical bursts explained by trapped fireballs ($\sim10 \mathrm{keV}$; \citealp{LinL20,Mereghetti20,LiCK21,Ridnaia21,Tavani21,Wada23}. 
The light curve of the XRB exhibits a multipeak structure, with the TOA showing some marginal deviation from the two peaks of FRB 200428 \citep{Mereghetti20,LiCK21,Ridnaia21}. Recently, \citet{Giri23} reanalyzed the CHIME radio data and provided a refined estimation of the TOAs for the two radio components, ruling out the simultaneity of the arrival times for the radio sub-bursts and corresponding X-ray peaks with a high significance. 
\citet{GeMY23} reconstructed the data of High Energy X-ray Telescope (HE) on board Insight-HXMT, and performed a detailed timing analysis by fitting the XRB light curve. Based on observations of INTEGRAL, Konus-Wind, and Insight-HXMT, the weighted average time delays of the two “main” X-ray peaks relative to the two radio peaks are found to be $3.47\pm0.73$ ms and $6.70\pm0.67$ ms, respectively \citep{Giri23}. In addition to the two narrow and bright “main” peaks, a weaker and softer peak with about half the magnitude and 20.55 ms prior to the radio signal is also noteworthy, where we refer to it as the “subpeak”. Moreover, INTEGRAL detected another narrow peak about 30 ms behind the second “main” peak \citep{Mereghetti20}. The timeline and interval of the X-ray peaks (two “main” peaks and a “subpeak”) and radio peaks (geocentric) discussed above are illustrated in Figure \ref{Fig1}. 

\subsection{Fitting Results of the FRB 200428-associated X-Ray counterpart}
\begin{figure}
	\centering
  	\subfigure[\label{Fig4a}]{
		\includegraphics[scale=0.62]{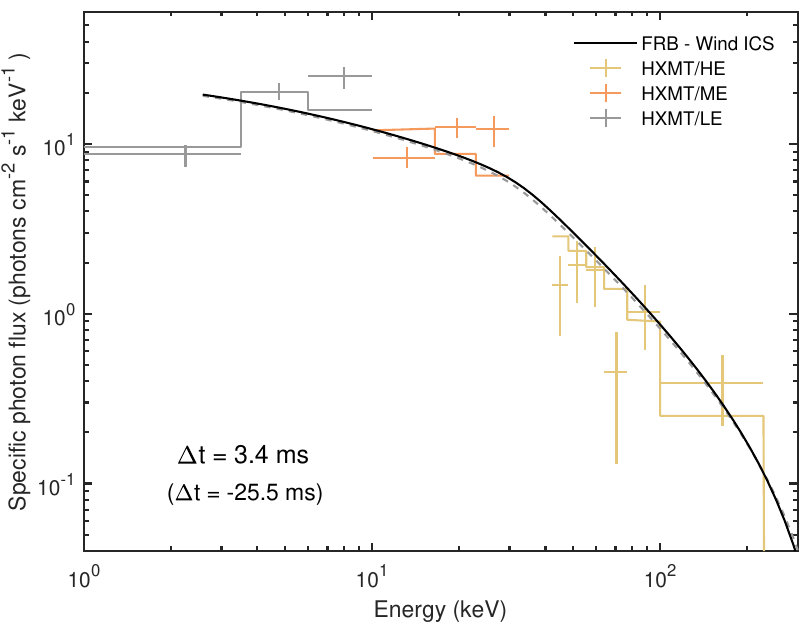}}
	\
	\subfigure[\label{Fig4b}]{
		\includegraphics[scale=0.62]{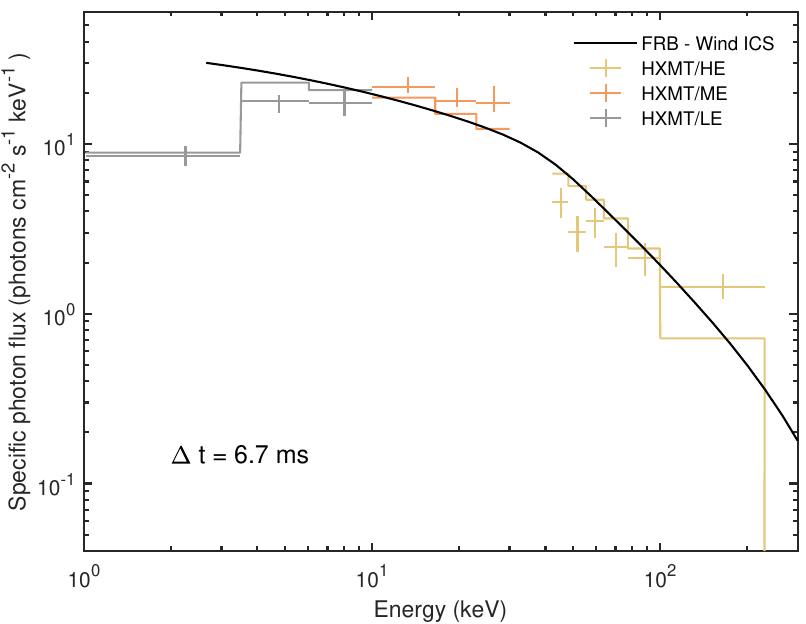} }
	\caption{X-ray spectrum of the FRB-outflow ICS model (black solid line) compared with the Insight-HXMT observation of the X-ray peaks (the grey, yellow and orange markers show the 1–10 keV, 5–30 keV, and 20–250 keV energy ranges, respectively; \citealp{GeMY23}). 
     (a) The first “main” X-ray peak (which could also be regarded as the precursor of the second radio peak, denoted by the dashed gray line). (b) The second  X-ray peak. }
	\label{Fig3} 
\end{figure}

We first analyze the two FRB peaks and the two “main” peaks of the associated XRB, taking into account the observed X-ray delays of 3.47$\,$ms and 6.76$\,$ms, respectively, as denoted by the red double-headed arrows in Figure \ref{Fig1}. 
The geometries and properties of the pair flow and FRB can be constrained by jointly fitting both the time lag and spectrum of the X-ray peak. 
In Figure \ref{Fig3}, we present the time delays and spectral energy distributions based on the FRB-outflow ICS model, in comparison with the Insight-HXMT observations of the X-ray peaks \citep{GeMY23} 
\footnote{It should be noticed that the observational data used here focus on the detailed spectra of the two X-ray peaks, with a few milliseconds width \citep{GeMY23}, rather than the entire 1.2-s spectrum in \citet{LiCK21}}.
The fiducial values of the fitting parameters are presented in Table \ref{Table1_2}. Here, we ignore the deviation in the low-energy part, due to the saturation and absorption effect, and include a constant to account for the saturation in the HE band.

\begin{table*}
\caption{The Parameters of the FRB-outflow ICS Model for the FRB 200428-associated X-Ray Peak}
\label{Table1_2}
\centering 
\setlength{\tabcolsep}{2.85mm}
\renewcommand{\arraystretch}{1.01}
\begin{threeparttable}
\begin{tabular}{cccccc} 
\hline
\hline
& & \multicolumn{2}{c}{\ Delayed  Peak \quad} & \multicolumn{2}{c}{Advanced Peak} \\
    \cmidrule{3-4} \cmidrule{5-6} 
& & First & Second  & Subpeak  & First  \tnote{a} \\
\hline
Scattering radius & $R_\mathrm{S}\,(R_\mathrm{L})$ & 1.5  & 1.5   & 1.5   & 1.5   \\
Incident angle & $\theta\,(10^{-1})$               & 1.3  & 1.6 & 1.3 & 1.3 \\ 
FRB swept angle &$\delta\theta\,(10^{-2})$         & 0.7  & 0.8 & 5.3 & 6.0 \\
FRB radiation cone angle &$\theta_\mathrm{frb}\,(10^{-1})$ & 1.2  &  1.5 & 0.8 & 0.7 \\
Bulk Lorentz factor &$\Gamma$    & 10 & 9 & - & 10  \\
Minimum Lorentz factor \tnote{b} &$\gamma^{\prime}_\mathrm{min}\,(10^4)$ & 5.0 & 3.5 & - & 3.5  \\
Maximum Lorentz factor \tnote{c} &$\gamma^{\prime}_\mathrm{max}\,(10^5)$ & 1.8 & 1.3 & - & 1.2  \\
Index of electrons distribution &$p$                              & 2 & 2 & - & 2  \\
Electron/positron luminosity &$L\,(10^{40}\mathrm{erg\,s^{-1}})$  & 2.8 & 4.0 & - & 3.6  \\
\hline
\end{tabular} 
\begin{tablenotes}
\item[a] ``First'' and ``Second'' represent the two “main” bright peaks. The advanced weaker ``Subpeak'' is not clearly quantified, and thus we here only constrain geometric parameters from the 20.55 ms advance.
\item[b] The prime denotes the quantity measured in the plasma comoving frame.
\item[c] We could only constrain the lower limit of $\gamma_\mathrm{max}^{\prime}$ due to the lack of higher-energy observations.
\end{tablenotes}
\end{threeparttable}
\end{table*}

The time shifts depend on the geometrical configuration, including the FRB swept angle $\mathrm{\delta}\theta$, the scattering angle $\theta$, the scattering radius $R_\mathrm{S}$, and the opening angle of the FRB emission cone $\theta_\mathrm{frb}$. However, these parameters are not independent, being related by $\theta_\mathrm{frb}=\theta-\delta\theta$ and $\theta\lesssim R_\mathrm{L}/R_\mathrm{w}$. Thus, the primary factors determining the time shift are the scattering angle $\theta$ and the FRB swept angle $\mathrm{\delta}\theta$, which influence the additional propagation length of the X-ray photons (the path difference) and the scattering phase relative to the LOS (the phase difference), respectively. 
As inferred from Equations \ref{eq:DopplerEIC} and \ref{eq:crosssection}, the shape of the spectrum is primarily determined by the product of the bulk and comoving frame Lorentz factors of the electrons, $\Gamma\gamma^\prime_\mathrm{c}$, and the scattering angle, $\theta$. 
While some degeneracy exists when considering the temporal or spectral features individually, this degeneracy of parameters could be significantly reduced if we jointly fit both features, allowing for a more robust interpretation of the model parameters in relation to the observable features.

On the other hand, if we associate the advanced and relatively weaker X-ray peak with the FRB as a paired event, then the X-ray peak could be regarded as the FRB precursor. As denoted by the blue double-headed arrows in Figure \ref{Fig1}, the new pairs would become then the first low-frequency CHIME burst and its weaker and softer X-ray peak 20.55 ms ahead, the second radio peak and its 25.55 ms precursor, leaving a separate X-ray peak. 
In this case, the time shift $\Delta t<0$, indicating that the FRBs are highly beamed.
For the X-ray subpeak and the first FRB peak, the geometry parameters are listed in the third row of Table \ref{Table1_2}, which can account for the 20 ms advance.
However, considering that the exact spectrum of the advanced X-ray peak is not clearly quantified, the properties of the electrons cannot be inferred, and the geometry angles are not well constrained due to the degeneracy.
For the first X-ray main peak and the second FRB peak, the spectrum and time shift can still be well fitted, as the parameters in the fourth row of Table \ref{Table1_2}.
As for the last X-ray peak, lacking a radio counterpart, as well as the third INTEGRAL peak, they may be explained by a premature cessation of FRB emission. In this case, the FRB radiation only interacted with the LOS relativistic electrons, then shut down before turning toward the observer, i.e. $\delta\theta<\theta-\theta_\mathrm{frb}$.

\subsection{X-Ray Precursor of FRB 221014}

\begin{figure}
	\centering
		\includegraphics[scale=0.63]{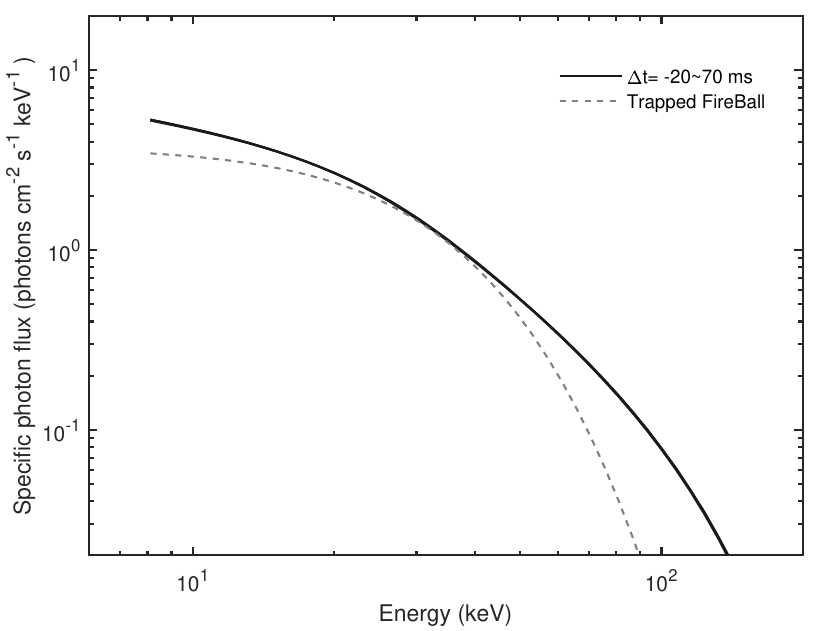} 
	\caption{ {Predicted spectral energy distribution of the FRB-outflow ICS model for the FRB 221014-associated X-ray counterpart, compared with the typical trapped fireball spectrum characterized by an effective temperature of $k_\mathrm{B}T_\mathrm{eff}\sim10\,\mathrm{keV}$. The maximum time advance (assuming the FRBs are highly beamed, i.e. $\delta\theta>>\theta_\mathrm{frb}$) ranges from 20 to 70 ms. The corresponding model parameters are listed in Table\ref{Table2}. We propose that in the FRB-outflow ICS model, the ICS process could generate the relatively harder X-ray spectra with different time advances.}
    }
    \label{Fig4} 
\end{figure}

FRB 221014 and its X-ray counterpart constitute the second FRB-XRB association event from SGR J1935+2154. However, since the burst was weak, the spectral index was poorly constrained by Knous-Wind's observation, given the low signal-to-noise ratio \citep{Frederiks22}. A refined analysis of the GECAM data, which offer higher temporal and spectral resolution, is ongoing and will be reported later \citep{WangCW22}.
In this work, we do not focus on the spectral fitting, due to the lack of detailed data. Instead, we propose that the ICS process of the FRB on an extreme pair flow could generate relatively harder X-ray spectra with different time advances. This is consistent with the higher $E_\mathrm{p}$ implied by Konus-Wind and the advance of the X-ray peak suggested by GECAM.
In order to compare with the forthcoming results of the data analysis, we constructively present the X-ray spectra predicted by the FRB-outflow ICS model with time advance from -20 to -70 ms. 

For the incident radio photons, we adopt a Gaussian-shaped FRB spectrum described by Equation \ref{eq:Gaussian}, with $\nu_\mathrm{p} = 1\, \mathrm{GHz}$, $\delta \nu = 500\, \mathrm{MHz}$ and isotropic luminosity $L_\mathrm{FRB} \sim 5\times 10^{36}\,\mathrm{erg\,s^{-1}}$. 
We consider an extreme scenario, in which the FRBs are highly beamed, such that the swept angle is much larger than its emission cone, i.e. $\delta\theta \gg \theta_\mathrm{frb}$ and $\theta \simeq \delta\theta$, in which case the advanced time is the longest.
As shown in Figure \ref{Fig4}, the FRB-outflow ICS model predicts a slightly harder spectrum with different advance times. The corresponding model parameters are listed in Table \ref{Table2}, which could be constrained more precisely with improved knowledge of the temporal and spectral properties of radio and X-ray emission.

\begin{table*}
\begin{center}
\caption{The Model Parameters of the FRB 221014-associated Advanced X-Ray Peak}
\label{Table2}
\setlength{\tabcolsep}{2.4mm}
\renewcommand{\arraystretch}{1.1}
\begin{threeparttable}
\begin{tabular}{cccccccc} 
\hline
\hline
$|\Delta t_\mathrm{max}|\, (\mathrm{ms})$    &20   & 30   & 40  & 50    & 60   & 70 \\ \hline
$R_\mathrm{S}\,(R_\mathrm{L})$    &   4.6   & 3.2  & 2.3 & 1.9   & 1.6  & 1.3\\
$\;\theta\simeq\delta\theta\;(10^{-1})$ & 0.4 & 0.6 & 0.8 & 1.1   & 1.3  & 1.5  \\ 

$\Gamma $          
                     &16 & 11 & 10  & 10  & 9 & 9\\
$\gamma_\mathrm{min}^{\prime}\,(10^4)$          
                     & 6.0 & 6.0 & 4.8 & 3.8 & 3.6 & 3.1\\
$\gamma_\mathrm{max}^{\prime}\,(10^5)$       
                    & 1.5& 1.5 & 1.2 & 1.0 & 0.9 & 0.8\\
$p$                 & 2.4& 2.4 & 2.5 & 2.5 & 2.6 & 2.7\\
$L (10^{40}\mathrm{erg\,s^{-1}})$    
                    & 63.0 & 15.0 & 4.0 & 1.7   & 0.8  & 0.5    \\
\hline
\end{tabular} 
\end{threeparttable}
\end{center}
\end{table*}

\section{Discussion} \label{sec:Discussion}

In this paper, we have proposed that the observed spectral and temporal features of FRB-associated XRBs can be attributed to the ICS of FRB photons by an extreme pair flow during magnetar activity. Further details and implications of the FRB–outflow ICS model are discussed as follows.

\textbf{1.\,\textit{Model predictions and potential observation evidence.}} \ 
Based on the FRB beaming characteristic and the geometry configurations of the model, various forms of FRB-XRB associations are possible. In some cases, only one of the radio or X-ray peaks may be detected, even though both are physically present: 
(1) if the FRB is triggered just before rotating to the LOS, resulting in a large scattering radius, or if the outflow is rather weak, the ICS emission may be too faint to detect--in this case, only the FRB would be observed; 
(2) if the FRB is quenched before rotating into the LOS, then only a bright hard-X-ray peak would be observed, but without an accompanying FRB detection; or
(3) a weaker radio burst may survive if the viewing angle is not far from the FRB emission cone, defined as a slow radio burst, characterized by a broader width and lower flux due to a smaller Doppler factor \citep{ZhangBing21,ChenConnery23}. 

Some observations potentially support this scenario. The multiple radio pulses with flux of a few janskys detected by the Yunnan 40m radio telescope on 2022 November 20 were associated with a multiple-peak-structured XRB detected by GECAM, which can be interpreted within this scenario (Li et al. 2024, in preparation). 
Moreover, additional radio bursts and their X-ray counterparts have been discovered from the same magnetar \citep{HuangYX22,LiXB22a,LiXB22b,Pearlman22,Younes22b}.
Beyond SGR 1935+2154, at least five bright XRBs from other magnetars also exhibit unusually hard spectra, which are clearly outliers among the hundreds of bursts detected by Konus-Wind since 1994 November (see \citealp{Ridnaia21} Extended Data Figure 6). Unfortunately, there were no simultaneous radio observations during these events, and thus they cannot be excluded as candidates for FRBs. These peculiar hard spectra could also be explained by the FRB-outflow ICS model, even in the absence of detected radio bursts. 

~

\textbf{2.\,\textit{Assumptions and limitations of the FRB-outflow ICS model.}} \
 In our model, we assume that FRBs are generated in the magnetosphere with a narrow beam that corotates with the magnetar, consistent with pulsar-like models. 
An increasing amount of observational evidence supports the origin of the magnetosphere: for instance, the polarization angle swings observed in some FRBs suggest a “pulsar-like” origin, in which the radiation is confined in a certain region of the dipole field \citep{LuoR2020,ZhangBing20a,ZhangBing23,LiuXH2025}, and the analysis of scintillation constrains the emission region of FRBs to be less than $\sim10^9\,\mathrm{cm}$ \citep{Nimmo2025}.  
Moreover, we further assume that the FRB emission cone rotates with the magnetar and persists for several to tens of milliseconds, corresponding to the intrinsic duration of the FRB emission, which is longer than the observed FRB duration. 
The persistent time of the FRB emission cone, i.e. the intrinsic duration of the FRB $\tau_\mathrm{frb}$, can be estimated from the swept angle, $\delta\theta$, $\tau_\mathrm{frb} = \delta\theta T/2\pi$. For the delayed X-ray peak, $\delta\theta\sim10^{-2}$, the intrinsic duration would be $\tau_\mathrm{frb}\sim 4\mathrm{ms}$, and if the X-ray peak is regarded to be advanced, $\delta\theta\sim5\times10^{-2}$ and $\tau_\mathrm{frb} \sim 20\mathrm{ms}$.

Meanwhile, a specific extreme pair flow is required to account for the significant hard ICS emission.
Despite the uncertainties and variability of the magnetar activity, we have proposed a plausible formation scenario for the extreme pair flow, potentially arising from the compression of the large-scale low-frequency pulse and violent reconnection in the current sheet (Section \ref{sec:formation}).
The parameter fitting results of the FRB 200428-associated X-ray peaks fall within a reasonable range.  
However, we should note that the violent activity of a magnetar and its highly deformed magnetosphere, where some closed field lines are opened, is quite complex and diverse. 
The corresponding particle acceleration processes could be much more complicated than those in typical steady-state neutron stars. Even in this case, the exact mechanism has not reached a consensus. 
On the one hand, theoretical MHD models suggest that $\Gamma_\mathrm{w}$ increases linearly until the FMS point, beyond which the acceleration slows down \citep{Beskin98,Contopoulos02}; alternatively, $\Gamma_\mathrm{w}$ may grow as $\propto r^{1/2}$ through reconnection in the stripped wind \citep{Lyubarsky01}. 
On the other hand, some models propose rapid acceleration within a narrow “acceleration region" to account for the sub-TeV emission of pulsars, although the exact mechanisms are not well understood \citep{Bogovalov00,Aharonian12}. And other models, such as the outer gap model \citep{Hirotani13} and the slot gap model \citep{Harding09}, have also been proposed to explain the high-energy emission of gamma-ray pulsars. Based on current observations, it is not easy to reach a consensus at present.
Considering these factors, other forms of electron motion resulting from different underlying mechanisms cannot be ruled out. We further discuss an ultrarelativistic wind scenario in the \hyperref[cold]{Appendix} for completeness, which may be applicable at least in the case of a young fast-rotating magnetar as the source of the FRB.

Furthermore, for a relatively young magnetar, which has undergone frequent active episodes, we assume that it is likely to posses a large inclination angle.
In a dipole field configuration, assuming that the FRB is emitted at a radius $\sim10^8\mathrm{cm}$, with its direction tangential to the local open field line \citep{LuWB20,ZhangBing22,LiuZN23}, the angle $\theta_\mu$ between the direction of the FRB and the magnetic axis can be expressed as $\tan \theta_\mu = 3\tan \theta_0 / (2-3\tan^2 \theta_0)$, where $\theta_0 \lesssim \theta_\mathrm{p} = \arcsin \sqrt{R_\mathrm{NS} / R_\mathrm{L}}$ denotes the polar angle (relative to the magnetic axis) of the emission point \citep{Qiao98}. 
Thus, while most of the energy of the extreme pair flow may be concentrated in the rotational equatorial zone, with the poloidal angle relative to the spin axis ranging from $\pi/2-\alpha $ to $\pi/2+\alpha$, 
the FRB could pass through this region and be upscattered, as long as the geometric condition $\alpha+\theta_\mu > \pi/2-\alpha$ is satisfied, suggesting an inclination angle $\alpha\gtrsim40^{\circ}$. 
\footnote{On the other hand, considering the presence of significant multipolar fields near the magnetar, if FRBs are triggered and emerge near the directions of multipolar magnetic poles at various latitudes \citep{YangYP21}, their emission could potentially intersect the equatorial zone, even if the inclination angle of the dipole field is small. However, the transparency of the FRB would be complex in such a scenario.}

In addition, given the uncertainties of the magnetar activity and the origin of FRBs, other scenarios, accounting for the FRB 200428-associated XRB remain possible, though they more or less overlook the temporal properties or do not fully explain why other XRBs lack such features. However, with only one well-analyzed association available so far, this is insufficient to definitively rule out or confirm these models. Nevertheless, as more radio and X-ray associations are detected, along with multiwavelength observations and statistical analyses, there will be greater promise for testing and constraining various models.

\section{Conclusions}   \label{sec:Conclusions}
The FRB 200428-associated XRB is significantly unique compared with other typical XRBs. It exhibits a harder spectrum, with a steeper power-law index and a much higher cutoff energy. Its light curve shows a multiple-peak structure, and the TOA of the XRB peak shows a time shift of a few millisecond relative to that of the FRB.
Similarly, FRB 221014 has also been found to have a slightly harder X-ray counterpart compared to other typical bursts in the same episode. 
These unique spectral and temporal properties of the FRB-associated XRB imply a special physical link between FRBs and their high-energy counterparts. 
In this paper, we have shown that both the hard spectrum and the peak time shift (whether delayed or advanced) of the X-ray peak can be explained by the ICS of FRB photons by an extreme pair flow at a scattering radius around the light cylinder (we refer to this scenario as the FRB-outflow ICS model). This pair flow is characterized by a bulk Lorentz factor $\Gamma\sim10$ and a power-law distribution in the comoving frame with a typical Lorentz factor $\gamma^\prime_\mathrm{m}\sim10^4$.
We propose that it could originate from the compression of a large-scale pulse due to rapid magnetophsere evolution and violent acceleration in the current sheet during magnetar activity.
The FRB-outflow ICS model fits well with the spectrum and time shift of FRB 200428 and its X-ray counterparts, with the parameters being in a reasonable range. From the fitting results, we can also infer the emission angle of the FRB $\theta_\mathrm{frb}\sim 10^{-1}$ and the swept angle $\delta \theta \sim 10^{-2}$, corresponding to an intrinsic duration of the radio emission approximating several to tens of milliseconds.
We constructively apply our model to the event of FRB 221014, providing a slightly harder spectrum, with time advances ranging from 20 to 70 ms. The model parameters for this event can be further constrained by future detailed analyses. 
The FRB-outflow ICS model also predicts various forms of association, which would be tested by other radio bursts and X-ray counterparts detected from SGR 1935+2154, as well as some XRBs exhibiting unusually hard spectra from different magnetars. 
We expect that with improvements in detector sensitivity and increasing numbers of simultaneous observations in radio and high-energy bands, more FRBs and associated hard XRBs will likely be discovered in the future, helping to reveal the origin and surrounding environment of FRBs.

\vspace{1em}
We thank the anonymous referee for providing helpful comments and suggestions that allowed us to improve our manuscript significantly.
We also thank Shao-Lin Xiong for his constructive suggestions and discussions about the measurements of the FRB-associated XRBs, and we thank F. A. Aharonian, D. Khangulyan, D. Frederiks, Xiao-Bo Li, Jin-Jun Geng, Jian-He Zheng, Ze-Nan Liu, Ken Chen, Jia Ren, Yu-Jia Wei, Zhen-yin Zhao and Bo-Yang Liu for their helpful discussions. Y. W. would like to thank De-kun Song for the insightful suggestions and helpful discussions.
This work was supported by the National Natural Science Foundation of China (grant Nos. 12393812 and 12273009) and the National SKA Program of China (grant Nos. 2020SKA0120302 and 2022SKA0130100). Y.P.Y. was supported by the National Natural Science Foundation of China grant No.12473047, the National Key Research and Development Program of China (2024YFA1611603) and the National SKA Program of China (2022SKA0130100), and acknowledges the support from the ``Science \& Technology Champion Project'' (202005AB160002) and from two ``Team Projects'' – the ``Innovation Team'' (202105AE160021) and the ``Top Team'' (202305AT350002), all funded by the ``Yunnan Revitalization Talent Support Program''.

\appendix
\section{ICS Process by a "Cold" Ultrarelativistic Wind}\label{cold}
The acceleration of particles has not reached a consensus yet. Other scenarios, such as rapid acceleration in narrow zones \citep{Bogovalov00,Aharonian12}, are also possible, though the exact mechanism remains unclear. 
Some works have proposed that electrons moving along the magnetic field line would be centrifugally accelerated if the inertia of the particles does not counteract the efficient corotation with the magnetic field. 
This can provide efficient electron acceleration to a high bulk Lorentz factor in some directions\citep{Osmanov09,Bogovalov14}.
For a young and fast-rotating magnetar, as the source of an FRB, the magnetocentrifugal mechanism could accelerate electrons to a bulk Lorentz factor of $\Gamma_\mathrm{w}\sim10^6 \tan\alpha \Lambda_{8}^{-1}P^{2}_{-2}B_{15}$, where $\Lambda$ is the number density in units of the Goldreich-Julian density and $\alpha$ is the angle between the magnetic field and the radial direction \citep{Osmanov09,Bogovalov14}. Thus, we refer to this as the “cold” ultrarelativistic wind.  
The direction of the accelerated electrons is along the tangent to the light cylinder surface, $\theta = (1-\Gamma_0/\Gamma_\mathrm{w})R_\mathrm{L}/r\simeq R_\mathrm{L}/r$ \citep{Aharonian03,Osmanov17}.
The anisotropic ICS of the directed FRB photons on the directed electron beam would produce a specific high-energy emission. Since particles in the magnetar wind move in a straight line with a high Lorentz factor, only the emission from electrons moving initially along the LOS contributes to observations. 
The spectrum can be described by
\begin{align}
    &\frac{dN_\mathrm{X}}{dE_\mathrm{X}dSdt}\left(E_\mathrm{X},\Gamma_\mathrm{w}\right)=\frac{\dot{N}_\mathrm{W}}{4\pi D_\mathrm{L}^2} 
    \iint \frac{ d \sigma \left[\Gamma_\mathrm{w}, \theta(r), E_\mathrm{X}, E_\mathrm{R}\right]}{dE_\mathrm{X}} \frac{dN_\mathrm{R}(r)}{dE_\mathrm{R}dS dr}dE_\mathrm{R}dl,
    \label{eq:dNdEdSdt}
\end{align}
where $E_\mathrm{X}$ and $E_\mathrm{R}$ represent the emission X-ray and incident radio photons, respectively, $\dot{N}_\mathrm{W}$ is the number rate of electrons/positrons in the magnetar wind, and $d N_\mathrm{R}(r) / d E_\mathrm{R} dS dr$ is the incident photon number at the scattering radius. For radio incident photons, $2E_\mathrm{R}\Gamma_\mathrm{w}\left[1-\cos \theta \right]\ll m_\mathrm{e}c^2$, which means that the Compton scattering takes place in the Thomson regime. The differential cross section of the anisotropic ICS in the Thomson regime is given by \citep{Fargion97,Dubus08} 
\begin{align}
    &\frac{ d \sigma \left[\Gamma_\mathrm{w}, \theta(r), E_\mathrm{X}, E_\mathrm{R}\right]}{dE_\mathrm{X}}= \frac{ \pi r_\mathrm{e}^2}{2 \beta \Gamma_\mathrm{w}^2E_\mathrm{R}} \left[3-\cos^2\theta’ +\frac{1}{\beta^2}\left(3 \cos^2\theta’-1\right)\left(\frac{E_\mathrm{X}}{\Gamma_\mathrm{w} E_\mathrm{R}’}-1\right)^2\right],
\end{align}
where the prime represents quantities in the comoving frame of the electrons. The time shift is also expressed by Eq. \ref{eq:deltat}. However, for a young magnetar as an FRB source at a cosmological distance, the millisecond-duration X-ray emission will be hard to detect.

~

\bibliography{main}{}
\bibliographystyle{aasjournal}

\end{document}